\documentstyle[eqsecnum,preprint,aps]{revtex}

\input epsf

\begin{document}

\preprint{DUKE-TH-96-129}

\draft

\title{Thermalization and Lyapunov Exponents\\ 
       in the Yang-Mills-Higgs Theory}

\author{
U. Heinz\cite{address}, C. R. Hu, S. Leupold\cite{address}, 
S. G. Matinyan\cite{address1}, and B. M\"uller}

\address{Department of Physics, Duke University, \\
Durham, North Carolina 27708--0305}

\maketitle

\begin{abstract}
  We investigate thermalization processes occurring at different time 
  scales in the Yang-Mills-Higgs system at high temperatures. We 
  determine the largest Lyapunov exponent associated with the gauge 
  fields and verify its relation to the perturbatively calculated 
  damping rate of a static gauge boson.
\end{abstract}


%%%%%%%%%%%%%%%%%%%%%%%%%%%%%%%%%%%%%%%%%%%%%%%%%%%%%%%%%%%%%%%%%%%%%%%%%%%
\section{Introduction}
\label{sec:intr}
%%%%%%%%%%%%%%%%%%%%%%%%%%%%%%%%%%%%%%%%%%%%%%%%%%%%%%%%%%%%%%%%%%%%%%%%%%%

To describe phenomena like heavy ion collisions or the electroweak phase 
transition in the Early Universe, the methods of quantum field theory 
must be applied to finite densities and temperatures. The most reliable 
approach, especially for the case of QCD, are lattice calculations in 
Euclidean space-time (for a recent overview see e.g.~\cite{Dehwa2}). 
Unfortunately the results are only applicable to equilibrium processes. 
Dynamical processes in quantum field theory, on the other hand, are most 
commonly studied using perturbation theory.
In theories with interacting massless degrees of freedom, however, 
naive perturbation theory in powers of the coupling constant breaks 
down at high temperature due to infrared divergences caused by the 
exchange of massless modes. To obtain finite results one must resum 
classes of Feynman diagrams which lead to dynamical mass generation and
thus regulate the infrared singularities. To do this systematically
one must disentangle the influence of different momentum scales;
in QCD at finite temperature, these are characterized by $T$, $gT$, and 
$g^2 T$ where the last two give the order of magnitude of the dynamical
masses for the originally massless electric and magnetic gluonic modes, 
respectively. An improved one-loop perturbation scheme based on these 
principles, called hard thermal loop resummation, was developed by 
Braaten and Pisarski \cite{BP90} (see also \cite{Thhwa2} for an overview). 
Using effective field theory methods, the above separation of the different 
momentum scales was achieved by Braaten and Nieto for static quantities 
at high temperature \cite{BN96}. 
 
At high temperature, it is 
reasonable to expect that phenomena which are 
dominated by long wavelength modes can be reliably described by 
classical field theory. This is the basic motivation for the work 
presented here where we study the classical dynamics of the coupled 
Yang-Mills-Higgs system. It is well known that the classical 
non-Abelian gauge field equations are nonintegrable and exhibit 
dynamical chaos \cite{BMM95}. Therefore one has to rely on numerical 
methods to study the evolution of classical fields. 

The real time evolution of classical SU($N$) Yang-Mills fields on 
a lattice, for $N=2,3$, was recently studied in Ref.~\cite{Bi94}.
Starting with randomly chosen initial conditions for the gauge fields 
on each lattice link it was found that the system shows chaotic behavior, 
i.e. the distance in phase space between two adjacent configurations 
increases exponentially in time. For systems with sufficiently weak 
coupling, the time-averaged growth rate (maximal Lyapunov exponent) 
$\lambda_{\rm N}$ was shown to scale linearly with the total energy 
of the system. The slope depends on the number $N$. 
Even if one starts with highly off-equilibrium configurations, the gauge 
fields thermalize very fast, i.e.~the energy distribution over the 
lattice soon exhibits a thermal shape. Therefore the total energy 
of the system can be related to the associated temperature, and the scaling
of the maximal Lyapunov exponent with the total energy can be reinterpreted
as a scaling with the temperature $T$ -- at least in the weak coupling 
regime. Precisely the relation 
 \begin{equation}
    \lambda_{\rm N} a \sim g^2 T a 
 \label{lyapscal}
 \end{equation}
which connects dimensionless quantities was established. Here $a$ is 
the lattice spacing and $g$ the coupling constant. One surmises that the 
actual value of the proportionality factor 
 \begin{equation}
   c_{\rm N} = {\lambda_{\rm N} a \over g^2 T a} 
 \end{equation}
survives in the continuum limit $a\to 0$. 

For the classical SU(2) and SU(3) gauge field systems, 
it was found in \cite{Bi94} 
%with high accuracy 
that the maximal 
Lyapunov exponent coincides numerically with twice the damping rate 
$\gamma_{\rm N}$ of a gauge boson at rest as calculated quantum field 
theoretically \cite{BP90} using thermal perturbation theory with hard 
thermal loop resummation:
 \begin{equation}
   {\lambda_{\rm N} a \over g^2 T a} = 2 {\gamma_{\rm N} \over g^2 T}  \,.  
 \label{scaling}
 \end{equation}
At first sight it is surprising that the Lyapunov exponent which is 
deduced from the classical field dynamics by an exact (i.e. 
non-perturbative) numerical simulation should be related at all
with the plasmon damping rate which is calculated perturbatively in 
quantum field theory. However, the close relation between linear response
functions calculated in the hard thermal loop approximation and classical 
transport phenomena has been known for many years \cite{Si60,He86}, 
indicating that the hard thermal loops are essentially classical
 (see also \cite{Ke94}). In addition it was argued in 
\cite{Bi95} that the damping of a gluon and the exponential growth of 
the distance between two adjacent classical trajectories are related. 
So far there is, however, no rigorous proof for the relation (\ref{scaling}). 

An empirical approach to the question whether the coincidence between the 
Lyapunov exponent and (twice) the damping rate is only accidental would be
to simulate other systems on the lattice and also calculate perturbatively 
their damping rates. A natural extension of the pure SU($N$) system is 
to couple it to a matter field. In this article we study the real-time 
evolution of the coupled SU(2) Yang-Mills-Higgs system. Again we are 
interested in the thermalization properties and the chaoticity of the 
fields. To compare with the high temperature results from hard thermal 
loop calculations we concentrate on the symmetric phase, i.e. we simulate 
the Higgs behavior in the electroweak plasma above the phase transition
in the $\theta_{\rm w}=0$ limit. Our work was partly motivated by
the recent study presented in Ref.~\cite{BT96} in which the question of 
the connection between the damping rate and the Lyapunov exponent in 
such a system was also addressed \cite{BT96}. This paper provides in
particular the analytical results for the plasmon damping rates for 
the gluon and Higgs fields in the hard thermal loop approximation.
However, in Ref.~\cite{BT96} the thermalization properties of the 
classical fields were not taken into account, and thus the determination 
of the temperature of the system was not dealt with in a satisfactory 
manner. The correct extraction of the temperature is, of course, crucial 
to establish relations like (\ref{lyapscal}). We postpone a more
detailed assessment of the conclusions of Ref.~\cite{BT96} until after
presenting our own results. 

The organization of the paper is as follows: In Sec.~\ref{sec:form} we 
formulate the problem. The thermalization properties of the fields 
are studied in Sec.~\ref{sec:therm}. In Sec.~\ref{sec:lyap} we compute 
the Lyapunov exponents, and we discuss our results and compare them with 
previous work in Sec.~\ref{sec:sum}.

%%%%%%%%%%%%%%%%%%%%%%%%%%%%%%%%%%%%%%%%%%%%%%%%%%%%%%%%%%%%%%%%%%%%%%%%%%%
\section{Formulation of the problem} 
\label{sec:form}
%%%%%%%%%%%%%%%%%%%%%%%%%%%%%%%%%%%%%%%%%%%%%%%%%%%%%%%%%%%%%%%%%%%%%%%%%%%

In this Section we present the realization of the coupled SU(2) 
Yang-Mills-Higgs system on the lattice, describe the initialization 
of the field variables, discuss the allowed ranges of the relevant 
parameters in the weak coupling and continuum limits, define the
Lyapunov exponents and describe how we measure them.

%%%%%%%%%%%%%%%%%%%%%%%%%%%%%%%%%%%%%%%%%%%%%%%%%%%%%%%%%%%%%%%%%%%%%%%%%%%
\subsection{The Yang-Mills-Higgs system on the lattice}   
\label{subsec:YMH}
%%%%%%%%%%%%%%%%%%%%%%%%%%%%%%%%%%%%%%%%%%%%%%%%%%%%%%%%%%%%%%%%%%%%%%%%%%%

In the temporal gauge $A_0^a=0$, the Hamiltonian of the coupled 
SU(2) Yang-Mills-Higgs system in the continuum is given by
 \begin{equation}
   H = \int d^3x 
     \left [ {\textstyle {1\over 2}} E^a_i E^a_i 
           + {\textstyle {1\over 2}} B^a_i B^a_i
           + \dot \Phi^{\dagger} \dot \Phi 
           + (D_i \Phi)^{\dagger} (D_i \Phi) 
           + \lambda (\Phi^{\dagger} \Phi)^2
     \right ]  \,.
 \label{H-Higgs}
 \end{equation}
The dot denotes the time derivative, and
 \begin{equation}
   \Phi = \left(  \begin{array}{c}
                     \phi_0-i \phi_1 \cr
                     \phi_2-i \phi_3 \cr
                   \end{array} 
          \right)
 \end{equation}
is a charged Higgs field in the fundamental representation of SU(2).
The gauge field is described by vector potentials $A_i^a(x)$. 
$E^a_i = -\dot A^a_i$ denote the electric fields, $B^a_i$ the magnetic 
fields, and $D_i$ the covariant derivatives. 

As mentioned before, the equations of motion derived from the Hamiltonian 
(\ref{H-Higgs}) are not integrable. One has to rely on numerical methods 
to study the time evolution of the fields. For numerical purposes we 
discretize the spatial coordinates by using a cubic lattice with $N^3$ sites.
Time is kept as a continuous variable. For details about the numerical 
simulation of the real-time evolution of the system we refer to \cite{Bi94}.

To realize the coupled SU(2) Yang-Mills-Higgs system on the lattice 
without losing its gauge symmetry properties we use link variables 
\cite{Wi74,Cr83} 
 \begin{equation}
   U_{x,i} = \exp \left( -\textstyle{{1\over 2}} iga A_i^c (x) \,\tau^c
                  \right)
 \label{eq7}
 \end{equation}
instead of vector potentials $A_i^c$. Here $\tau^c$ are the Pauli matrices. 
The covariant derivative acting on the Higgs field is replaced by the 
difference
 \begin{equation}
   {1 \over a} \left( U_{x,i} \,\Phi_{x+i} - \Phi_x \right) \,.
 \end{equation}
The electric field strength on the lattice is related to the link variables
and their time derivatives by 
 \begin{equation}
   E^a_{x,i} = - \frac{i}{g a} {\rm tr}
   (\tau^a \dot U_{x,i} \, U^{\dagger}_{x,i}) \,.
 \label{EUdot}
 \end{equation}
The magnetic energy is expressed in terms of the so-called plaquette 
operator $U_{x, ij}$, which is the product of all four link variables 
on an elementary plaquette with sites $(x, x+i,x+i+j,x+j)$:
 \begin{equation}
   U_{x,ij} = U_{x,i} \, U_{x+i,j} \, U_{x+i+j,-i} \, U_{x+j,-j}
 \end{equation}
with $U_{x,-i} = U^{\dagger}_{x-i,i}$. The links are directed and hence 
the plaquettes are oriented. In the continuum limit the plaquette variable 
$U_{x,ij}$ is related to the local magnetic field $B_{x,k}^a$ by
 \begin{equation}
   U_{x,ij} = \exp\left(
            - \textstyle{{1\over 2}} i g a^2 \epsilon_{ijk} B^a_{x,k} \tau^a
                  \right) \,. 
 \label{eqB}
 \end{equation}
It is useful to write the SU(2) matrices in the form
 \begin{equation}
   U = u_0 - i \tau_a u_a 
     = \left(  \begin{array}{rr}
                   u_0-i u_3\,, & -u_2 - i u_1 \cr
                   u_2-i u_1\,, &  u_0 + i u_3 \cr
                \end{array} 
       \right)
 \label{Uquat}
 \end{equation}
where the $u_i$ are four real numbers, which can be thought of as components
of a quaternion. Since $\det U=1$, each link variable $U$ can be expressed in 
terms of three angles, $\rho_{\rm G}$, $\vartheta_{\rm G}$, $\varphi_{\rm G}$,
characterizing a three dimensional hypersphere:
 \begin{equation}
   U = \cos {\rho_{\rm G}\over 2} 
   - i \hat \rho_{\rm G}^a \tau^a \sin {\rho_{\rm G}\over 2} 
 \label{rhoG}  
 \end{equation}
with a three dimensional unit vector 
 \begin{equation}
   \hat \rho_{\rm G} 
   = \left( \sin\vartheta_{\rm G} \cos\varphi_{\rm G},
            \sin\vartheta_{\rm G} \sin\varphi_{\rm G},
            \cos\vartheta_{\rm G}
     \right) \,.
 \end{equation}
Similarly the four independent components of the Higgs field can be 
represented as a quaternion:
 \begin{equation}
   \Phi = \phi_0 - i \tau_a \phi_a = 
   \left( \begin{array}{rr}
            \phi_0-i\phi_3 \,, & -\phi_2-i\phi_1 \cr
            \phi_2-i\phi_1 \,, &  \phi_0+i\phi_3 \cr
          \end{array} 
   \right)  \,.
 \end{equation}
In contrast to the link variable $U$ the Higgs quaternion $\Phi$ has 
arbitrary length $R$ defined by
 \begin{equation}
   R^2 = \phi_0^2 + \phi_1^2 + \phi_2^2 + \phi_3^2 
       = {\textstyle {1\over 2}} \, {\rm tr} \left(\Phi^\dagger \Phi\right)
   \,.  
 \label{defHiggsamp}
 \end{equation}
Again we can introduce three angles to get the representation 
 \begin{equation}
   \Phi = R \left( \cos {\rho_{\rm H}\over 2} 
               - i \hat \rho_{\rm H}^a \tau^a \sin {\rho_{\rm H}\over 2} 
            \right)  \,.
 \end{equation}
Using the quaternion representations for the link variables and the Higgs 
fields, the Hamiltonian of the coupled SU(2) Yang-Mills-Higgs system on the 
lattice reads 
 \begin{equation}
   H = H_{\rm el} + H_{\rm mag} + H_{\rm kin} + H_{\rm pot} \,,
 \label{Htot}
 \end{equation}
where we define the electric energy
 \begin{equation}
   H_{\rm el} = \sum\limits_{x} \epsilon_{\rm el}(x)  \,, 
 \label{Hel}
 \end{equation}
the magnetic energy
 \begin{equation}
   H_{\rm mag} = \sum\limits_{x,i,j} \epsilon_{\rm mag}(x,ij)   \,, 
 \label{Hmag}  
 \end{equation}
the Higgs kinetic energy
 \begin{equation}
   H_{\rm kin} = \sum\limits_{x} \epsilon_{\rm kin}(x)  \,, 
 \label{Hkin}
 \end{equation}
and the Higgs potential energy
 \begin{equation}
   H_{\rm pot} = \sum\limits_{x} \epsilon_{\rm  pot,2}(x) 
               + \sum\limits_x \epsilon_{\rm  pot,4}(x) 
               + \sum\limits_{x,i} \epsilon_{\rm  int}(x,i) \,,  
 \label{Hpot}
 \end{equation}
and the $\epsilon$'s are energies associated with the respective site, link,
or plaquette:
 \begin{mathletters}
 \label{epsilon}
 \begin{eqnarray}
   \epsilon_{\rm el}(x) 
   &=& a^3 \sum\limits_{i}{\textstyle {1\over 2}} \, E_{x,i}^a E_{x,i}^a \,,  
 \label{Eel}\\
   \epsilon_{\rm mag}(x,ij) 
   &=& {4 \over g^2 a} \left[ 1-{\textstyle {1\over 2}}\, {\rm tr}\,U_{x,ij}
                       \right] \,, 
 \label{Emag}\\
   \epsilon_{\rm kin}(x) 
   &=& a^3 \left[ {\textstyle {1\over 2}} \, {\rm tr} 
           \left( \dot \Phi_x^\dagger \dot \Phi_x  
           \right)
           \right]  \,, 
 \label{Ekin}\\
   \epsilon_{\rm  pot,2}(x) 
   &=& 6a \left[{\textstyle {1\over 2}} \,{\rm tr}(\Phi_x^\dagger \Phi_x)
          \right] \,, 
 \label{Ephi}\\
   \epsilon_{\rm  pot,4}(x)
   &=& a^3 \lambda \left[ {\textstyle {1\over 2}}\, {\rm tr} 
                          \left( \Phi_x^\dagger \Phi_x \right) 
                   \right]^2  \,, \
 \label{Elambda}\\
   \epsilon_{\rm  int}(x,i)
   &=& - a \,{\rm tr} \left(\Phi_x^\dagger U_{x,i} \Phi_{x+i}\right) \,.
 \label{Egh} 
 \end{eqnarray}
 \end{mathletters}
To derive the last equation we have exploited the unitarity of the link 
variable $U$ and the fact that the trace of a quaternion is real.

The Hamiltonian (\ref{Htot}) is scale invariant. To see this we define the 
dimensionless variables
 \begin{equation}
   \bar t = t/a  \,, \quad \bar H = g^2 a H \,, \quad 
   \bar E_{x,i}^a = g a^2 E_{x,i}^a \,, \quad 
   \bar \Phi_x = ag\,\Phi_x  \,, \quad 
   \mbox{and} \quad 
   \bar\lambda = \lambda /g^2  \,.
 \end{equation}
In terms of these we get
 \begin{eqnarray}
   \bar H &=& 
   \sum\limits_{x,i} {\textstyle {1\over 2}} \bar E_{x,i}^a \bar E_{x,i}^a
   + 4 \sum\limits_{x,i,j} 
       \left[1-{\textstyle {1\over 2}} \, {\rm tr} \,U_{x,ij}\right]
 \label{Hscal}\\ 
   &+& \sum\limits_{x} {\textstyle {1\over 2}} \, 
       {\rm tr} \left( \dot {\bar\Phi}\vphantom{\bar\Phi}_x^\dagger 
                       \dot {\bar \Phi}_x  
                \right)
     + \sum\limits_{x,i} 
       {\rm tr} \left[  \bar\Phi_x^\dagger \bar\Phi_x 
                      - \bar\Phi_x^\dagger U_{x,i} \bar\Phi_{x+i} 
                \right]
     + \bar\lambda \sum\limits_x {\textstyle {1\over 4}}
                \left[ {\rm tr} \left( \bar\Phi_x^\dagger \bar\Phi_x \right) 
                \right]^2
 \nonumber
 \end{eqnarray}
where the dot now denotes the derivative with respect to the scaled time
$\bar t$. Therefore the only parameters in the system are the scaled energies
and the scaled Higgs self coupling $\bar\lambda$ while the lattice spacing and
the gauge coupling constant do not appear explicitly.  

It is easy to derive the equations of motion from the lattice Hamiltonian
(\ref{Hscal}) and the relations (\ref{EUdot}) and (\ref{Uquat}): 
 \begin{mathletters}
 \label{EOM}
 \begin{eqnarray}
   (\dot u_0)_{x,i} 
   &=& {\textstyle \frac{1}{2}} \bar E^a_{x,i} u^a_{x,i} \, ,
 \label {U0dot}\\
   (\dot u_a)_{x,i} 
   &=& {\textstyle -\frac{1}{2}} 
       \left [ \bar E^a_{x,i} (u_0)_{x,i} + \epsilon^{abc} 
               \bar E^b_{x,i} (u_c)_{x,i}
       \right] \,,
 \label{Udot}\\
   \dot {\bar E}\vphantom{E}^a_{x,i} 
   &=& i \sum\limits_j  \, {\rm tr} \left( \tau^a\,U_{x,ij} \right) 
       + {\bar j}^a_{x,i}  \,, 
 \label{eomE}\\
   \ddot {\bar \Phi}_x 
   &=& - \bar\lambda \, {\rm tr} \left(\bar\Phi_x^\dagger \bar\Phi_x \right) 
         \bar\Phi_x - 6 \, \bar\Phi_x 
       + \sum\limits_i \left( U_{x,i}\, \bar\Phi_{x+i} 
                            + U_{x-i,i}^\dagger \bar\Phi_{x-i} \right)  \,. 
 \label{Phiddot}
 \end{eqnarray}
 \end{mathletters}
Here
 \begin{equation}
   {\bar j}^a_{x,i}={\textstyle {i\over 2}} \, {\rm tr} 
   \left( \bar\Phi_x^\dagger \tau^a U_{x,i}\, \bar \Phi_{x+i}  
   \right)  
 \label{hj}
 \end{equation}
is the lattice version of the gauge-covariant isospin current of the 
Higgs field.

In addition to satisfying these equations of motion the physical solutions
must obey Gauss'\ law\footnote{For brevity we present only the 
continuum version of Gauss' law.}
 \begin{equation}
   D_i^{ab} E_i^b - {\textstyle {1\over 2}} ig 
   \left( \Phi^\dagger \tau^a \dot \Phi - \dot \Phi^\dagger \tau^a \Phi 
   \right) = 0 
 \label{gauss}
 \end{equation}
with the covariant derivative $D_i^{ab}$ in the adjoint representation. 
The expression on the l.h.s. of (\ref{gauss}) is a conserved quantity since 
it commutes with the Hamiltonian of the system. Thus to satisfy Gauss'\ law 
for all times we must only make sure that our initial conditions obey 
(\ref{gauss}). 

For a study of the thermalization properties of the different degrees 
of freedom the energy transfer between different sectors of the system
is an important quantity. As we will see below the energy transfer from 
the gauge part of the system to the Higgs part is of particular interest. 
It is given by
 \begin{equation}
   {d \over d\bar t} \left( \bar H_{\rm el}+ \bar H_{\rm mag} \right)
   = - {d \over d\bar t} \left( \bar H_{\rm kin}+ \bar H_{\rm pot} \right) 
   = \sum\limits_{x,i} \bar E_{x,i}^a \, {\bar j}_{x,i}^a  
 \label{rate}
 \end{equation}
as can be easily derived from the equations of motion. The local Higgs 
isospin current ${\bar j}_{x,i}^a$ is defined in (\ref{hj}).

%%%%%%%%%%%%%%%%%%%%%%%%%%%%%%%%%%%%%%%%%%%%%%%%%%%%%%%%%%%%%%%%%%%%%%%%%%%
\subsection{Initialization, weak coupling and continuum limit}  
\label{subsec:iwc} 
%%%%%%%%%%%%%%%%%%%%%%%%%%%%%%%%%%%%%%%%%%%%%%%%%%%%%%%%%%%%%%%%%%%%%%%%%%%

We initialize the system in the same way as in \cite{Bi94} by choosing 
random initial configurations:

\begin{itemize}

\item The isospin directions of the Higgs fields, characterized by the angle 
variables $\rho_{\rm H}$, $\vartheta_{\rm H}$, $\varphi_{\rm H}$ are chosen 
randomly for each lattice site. To keep control over the 
initial potential energy of the Higgs fields the length of each Higgs 
quaternion is {\it not} randomly chosen. Instead we take for simplicity the 
same length $R$ for all $\Phi_x$ initially. 
Obviously the initial Higgs potential energy (\ref{Hpot}) increases with $R$. 

\item The isospin directions of the gauge fields are also randomly chosen. 
In this case, however, this does not mean that {\em all} the angles are 
arbitrary. The reason is that not only the isospin direction of the 
vector potential $A$ but also its amplitude enters the link variable 
$U$ as an angle variable. It is easy to see that $\rho_{\rm G}$ is 
connected to the amplitude while $\vartheta_{\rm G}$ and $\varphi_{\rm G}$ 
characterize the isospin direction. Therefore the latter two
angles are initialized with random values on each link. To limit
the initial magnetic energy we restrict the gauge field amplitude. Thus we 
introduce a parameter $\delta$ which controls the initial magnetic energy 
and choose $\rho_{\rm G}$ on each link randomly but within the 
range $[0,2\pi\delta]$. For $\delta =0$ the initial magnetic energy 
also vanishes. For small $\delta$ the relation $H_{\rm mag} \sim 
\delta^2$ holds. 

\item To satisfy Gauss' law (\ref{gauss}) in an easy way we choose 
initially vanishing field momenta on each link, i.e.~$E = 0 = \dot\Phi$. 
This implies that the initial electric energy and the Higgs kinetic 
energy, (\ref{Hel}) and (\ref{Hkin}) respectively, vanish. Note that 
with non-vanishing potential but vanishing kinetic energies the system 
is initially in a highly off-equilibrium state. 

\end{itemize}

When initialized in this way the system's further evolution is 
determined by the three parameters $\delta$, $R$, and $\bar\lambda$, 
or by the initial magnetic energy and the initial Higgs potential 
energy (quadratic and quartic part), respectively. 

To make contact with perturbative calculations it is necessary to stay 
in the weak coupling regime. Furthermore, weak coupling is required 
for the classical calculation to be a valid approximation to the full
quantum treatment. This puts constraints on the range of the physical 
parameters $\delta, R, \bar\lambda$. The scale invariance of the 
Hamiltonian allows to fix the gauge coupling at $g=1$ and the lattice 
spacing at $a=1$.

Weak coupling within the gauge sector is realized if the link variables 
are in the vicinity of unity (cf.~(\ref{eq7})). In the initial state this 
is achieved by choosing $\delta$ small which implies that $\rho_{\rm G}$ 
in (\ref{rhoG}) is small. On the other hand, as first noted in \cite{MT92} 
and recently worked out in \cite{Mu96}, one encounters finite size 
artifacts in the pure SU(2) system when using too small values of 
$\delta$ on a finite lattice. To explore the weak gauge coupling
regime one is thus restricted to a window of not too small and not too large 
values for $\delta$, or one must take very large lattices. With the latter
option one soon runs into computer limitations.

From the Hamiltonian (\ref{Hscal}) one sees that weak coupling within 
the Higgs sector is ensured if the quadratic terms dominate the quartic 
one. A system with weak Higgs self coupling can thus be initialized 
by requiring 
 \begin{equation}
   \lambda R^4 < R^2  \,. 
 \label{weakcoupH} 
 \end{equation}

To ensure weak coupling between the gauge and the Higgs fields, the 
Higgs isospin current $j$ in (\ref{eomE}) should be small compared 
to the magnetic energy contribution\footnote{In the Abelian and 
continuum limits the first contribution on the r.h.s.~of Eq.~(\ref{eomE}) 
is $\nabla\times {\bf B}$.}. Estimating both contributions by the 
(maximal) field amplitudes we get the condition
 \begin{equation}
   R^2 < \delta  \,. 
 \end{equation}

The continuum limit also puts constraints on the possible parameter ranges.
First, as seen from Eq.~(\ref{eq7}), to recover the continuum limit
from the lattice formulation one needs $gaA(x)\to 0$. If the lattice 
simulations are supposed to yield a reasonable approximation to continuum 
behaviour, $\delta$ must therefore not be large. Another lattice artifact 
shows up if the total energy in the system is too high. Due to the 
compactness of the space of link variables the magnetic energy on the 
lattice (\ref{Hmag}) is limited from above by
 \begin{equation}
   (\bar H_{\rm mag})_{\rm max} = 4 \sum\limits_{x,i,j} 
   [1-{\textstyle {1\over 2}} \, {\rm tr} (-1)]  = 24 N^3  \,. 
 \label{magmax} 
 \end{equation}
If the total energy in the system is much larger than this value,  
equipartition of the energy among the magnetic, electric and Higgs 
degrees of freedom may thus not be possible.

All these considerations must be taken into account when selecting 
the initial parameters $\delta$ and $R$ as well as the value $\bar \lambda$ for
the Higgs self-coupling. In our calculations we have made sure to
choose these parameters in such a way that we are simultaneously in the 
weak coupling regime and close to the continuum limit.

%%%%%%%%%%%%%%%%%%%%%%%%%%%%%%%%%%%%%%%%%%%%%%%%%%%%%%%%%%%%%%%%%%%%%%%%%%%
\subsection{Lyapunov exponents}
\label{subsec:lyap}
%%%%%%%%%%%%%%%%%%%%%%%%%%%%%%%%%%%%%%%%%%%%%%%%%%%%%%%%%%%%%%%%%%%%%%%%%%%

In the next Section we will study the thermalization properties of systems
initialized as described above. The self-thermalization of the fields
is strongly influenced by the chaoticity in its dynamics which is
characterized by a set of positive Lyapunov exponents. We will here
concentrate on the largest Lyapunov exponent associated with the 
most unstable dynamical mode. In principle one can define two such
exponents, $\lambda_{\rm G}$ and $\lambda_{\rm H}$, for the gauge and 
Higgs fields, respectively, by using two different distance measures in
the space of field configurations. The first measures the growth rate of 
the Euclidean distance in the gauge field sector,
 \begin{equation}
   D^2_{\rm G}[U,U'] = \sum\limits_{x,i} \sum\limits_{\alpha =0}^3 
   \left[ ( (u_\alpha)_{x,i} - (u'_\alpha)_{x,i} )^2  
   +( (\dot u_\alpha)_{x,i} - (\dot u'_\alpha)_{x,i} )^2 \right] \, ,
 \label{Gdist}
 \end{equation}
where $U$ and $U'$ denote the gauge fields associated with two solutions 
of the equations of motion (\ref{EOM}) resulting from two initially very 
close field configurations. The other Lyapunov exponent measures similarly 
the growth rate in the Higgs sector,
 \begin{equation}
  D^2_{\rm H}[\Phi,\Phi'] = \sum\limits_{x} 
  \sum\limits_{\alpha =0}^3 
  \left[ ( (\phi_\alpha)_{x} - (\phi'_\alpha)_{x} )^2 
  +( (\dot \phi_\alpha)_{x} - (\dot \phi'_\alpha)_{x} )^2 \right]  \,,
 \label{Hdist}
 \end{equation}
where $\Phi$ and $\Phi'$ are the corresponding Higgs fields from the same
two solutions. As we will see in the following, the two Lyapunov exponents 
$\lambda_{\rm G}$ and $\lambda_{\rm H}$ turn out to be identical. 

As usual the Lyapunov exponent is defined as
 \begin{equation}
   \lambda = \lim_{t\rightarrow\infty}\lim_{D(0)\rightarrow 0}
   {1\over t} \ln {D(t)\over D(0)}  \,.   
 \label{lyap_def}
 \end{equation}
To avoid the problem of distance saturation \cite{Bi94} in directly 
measuring the Lyapunov exponent, we use the rescaling method 
for the evaluation of the large-time limit required in (\ref{lyap_def}) 
(see \cite{Bi94} for more details). 

In principle other definitions for the distance between field 
configurations are possible. In particular, one might wish to replace 
the Higgs distance (\ref{Hdist}) by a gauge invariant definition like e.g.
\cite{Bi94,BT96}
 \begin{equation}
   \sum\limits_x \vert R^2_x - {R_x'}^2 \vert \,.
 \label{gidist}
 \end{equation}
However, the Euclidean distances defined above are more appropriate for 
the rescaling method. For the pure SU(2) gauge system it was carefully
checked that different distance measures give rise to the same Lyapunov 
exponent \cite{Go94}.

%%%%%%%%%%%%%%%%%%%%%%%%%%%%%%%%%%%%%%%%%%%%%%%%%%%%%%%%%%%%%%%%%%%%%%%%%%%
\section{Thermalization processes}  
\label{sec:therm}
%%%%%%%%%%%%%%%%%%%%%%%%%%%%%%%%%%%%%%%%%%%%%%%%%%%%%%%%%%%%%%%%%%%%%%%%%%%

In this Section we investigate various thermalization processes in the 
coupled SU(2) Yang-Mills-Higgs system. We will encounter different types of
thermalization processes which operate on different time scales. 
In particular we find that, starting with the initialization described 
in Sec.~\ref{sec:form}, the gauge and the Higgs subsectors each thermalize
rather rapidly, while the equipartition of energy between them requires a 
much longer time. In the following two subsections we first present 
results on the fast thermalization processes associated with the gauge and 
the Higgs subsystems, and then discuss the system's long-time behaviour.
For reasons given in Sec.~\ref{sec:form} our study is confined to 
certain parameter ranges which are consistent with weak coupling and
proximity to the continuum limit.

%%%%%%%%%%%%%%%%%%%%%%%%%%%%%%%%%%%%%%%%%%%%%%%%%%%%%%%%%%%%%%%%%%%%%%%%%%%
\subsection{Short-time thermalization and relaxation times}
\label{subsec:short}
%%%%%%%%%%%%%%%%%%%%%%%%%%%%%%%%%%%%%%%%%%%%%%%%%%%%%%%%%%%%%%%%%%%%%%%%%%%

%%  {\bf Dear Chao-ran, the following paragraph is completely opaque to me.
%%  I just don't understand what you want to say, so I didn't try rewriting it.
%%  Please discuss its formulation with Berndt. Ulrich} 

Thermalization processes in a large system can be studied by following the
evolution of the energy distribution over the microscopic degrees 
of freedom. In general, the distribution should approach that of a Gibbs 
ensemble upon equilibrium. As shown in the cases of pure SU(2) and SU(3) 
gauge fields \cite{Bi94}, by measuring appropriate energy distributions, one can 
determine a parameter $T_{\rm a}$ which we call the {\it apparent 
temperature} henceforth. The apparent temperature $T_{\rm a}$ is
related to the system's true temperature $T$, but generally differs from $T$,
if the degrees of freedom of the system are not all independent, such as for a
gauge theory.

We will test the degree of thermalization in various subsectors of the 
system by comparing the distributions for the associated energies over 
the lattice with a Boltzmann distribution
 \begin{equation}
   P(\epsilon)=N_0 f(\epsilon) \exp(-\epsilon/T_{\rm a}) \,.
 \label{distri}
 \end{equation}
Here $\epsilon$ is the energy associated with the selected
degrees of freedom (e.g. the magnetic or electric energy of the gauge fields),
and $f(\epsilon)$ is an appropriate phase space factor, specified below.
$N_0$ is a normalization constant. The distribution 
$P(\epsilon)$ is obtained in the simulation by sampling, at a fixed time $t$,
the variable $\epsilon$ over all sites or plaquettes on the lattice; 
it gives
the probability, averaged over the whole lattice, to find the value $\epsilon$
at a given, though arbitrary, lattice point. If this distribution
has the Boltzmann shape (\ref{distri}) the system has lost all initial order
in the distribution of $\epsilon$, and the entropy has reached a maximum
under the constraint of fixed total energy associated with the selected 
degrees of freedom.

We will present measurements for four different energy distributions,
the distribution $P_{\rm el}(\epsilon)$ for the electric energy
$\epsilon_{\rm el}$ of the gauge fields, the distribution 
$P_{\rm mag}(\epsilon)$ of the magnetic energy $\epsilon_{\rm mag}$, 
as well as the distributions $P_{\rm kin}(\epsilon)$ and 
$P_{\rm pot,2}(\epsilon)$ of the kinetic and potential energies
of the Higgs field, $\epsilon_{\rm kin}$ and $\epsilon_{\rm  pot,2}$, 
respectively, as defined in Eqs.~(\ref{epsilon}). In the weak coupling 
limit the respective phase space factors $f(\epsilon)$ are given by
 \begin{equation}
   f(\epsilon)=
   \cases{\epsilon^2 &for $P_{\rm el}(\epsilon)$, \cr
          \sqrt{\epsilon(8-\epsilon)} &for $P_{\rm mag}(\epsilon)$, \cr
          \epsilon &for $P_{\rm kin}(\epsilon)$, \cr
          \epsilon &for $P_{\rm pot,2}(\epsilon)$.\cr}
 \label{factor}
 \end{equation}
A detailed calculation of the phase space factors can be found in  
Appendix \ref{sec:appendix}.

By plotting $\ln[P(\epsilon)/f(\epsilon)]$ against $\epsilon$ one can 
read off the apparent temperature $T_{\rm a}$. It was discussed in 
\cite{Bi94,Thesis94} that $T_{\rm a}$ is not in all cases identical
with the true temperature of the system. Writing
 \begin{equation}
   T_{\rm a} = C(T) \, T \,,
 \label{Tcorr}
 \end{equation}
one finds
 \begin{equation}
   C = \cases{1 &for $P_{\rm el}(\epsilon)$, \cr
              C_{\rm m} &for $P_{\rm mag}(\epsilon)$, \cr
              1 &for $P_{\rm kin}(\epsilon)$, \cr
              1 &for $P_{\rm pot,2}(\epsilon)$. \cr}
 \label{Ccorr}
 \end{equation}
$C_{\rm m}$ is a number varying between $2/3$ and $1$ as $T$ rises from 
low to high values \cite{Bi94}. In Ref.~\cite{Bi94} it was argued that
this can be understood in terms of the changing number of effective 
degrees of freedom which at low temperatures is only $2/3$ of that at 
high temperatures, due to the increasing importance of longitudinal 
plasma modes at high temperature. More details on the relation between
the apparent temperature and the actual temperature can be found in
the Appendix \ref{sec:appendix}. 

By numerically evolving the initial configurations forward in time, 
we can thus follow the system's thermalization. At each moment, we 
can measure its energy distributions and then read off its characteristic 
temperatures from the measured distributions, using (\ref{Tcorr}) for 
correct normalization. In Fig.~\ref{F1} several characteristic temperatures 
of the system are plotted as functions of time, for short times 
below $t=6$ (in lattice units). To ensure weak coupling and closeness to the
continuum limit we choose $\delta=0.2$, $R=0.2$, and $\lambda=1$. 
In this case it was found that $C_{\rm m}\approx 0.687$.

The curves in Fig.~\ref{F1} demonstrate a fast thermal equilibration 
within the gauge field sector between the electric (solid line) and 
magnetic (dashed line) energy distributions. After some transient oscillations
they approach, on a typical time scale of 1 lattice unit, a common 
temperature $T_{\rm el} \approx T_{\rm mag} \approx 0.47$.
On roughly the same time scale the kinetic (dotted line) and quadratic 
potential (dot-dashed line) parts of the Higgs field energies also equilibrate,
but at a much lower temperature $T_{\rm kin} \approx T_{\rm pot,2} 
\approx 0.07$. This indicates that in the coupled SU(2) Higgs system 
the gauge and the Higgs subsector each undergo a separate, fast 
equilibration, starting from a highly off-equilibrium initial state. 

The approximate equality of the two associated relaxation times, 
$\tau_{\rm G}$ for the gauge sector and $\tau_{\rm H}$ for the Higgs sector,
can be easily understood if one makes the following two hypotheses: 
\newline
1. Thermalization in a chaotic system is driven by the most unstable modes
and thus dominated by its largest Lyapunov exponent. 
\newline
2. The gauge and the Higgs sectors are associated with the same 
largest Lyapunov exponent. 
\newline
While it is difficult to find a rigorous proof for the first hypothesis, 
such a behaviour is intuitively expected. The second hypothesis
can be checked in our simulation and is found to be correct
(see Sec.~\ref{sec:lyap}). 

At short times, each subsector settles down at a common temperature
for all its degrees of freedom, but the temperatures of the two 
subsectors differ widely. This can be traced to the different initial
energy content in the two subsectors. On average, the energy per degree
of freedom in the gauge sector is initially 0.478, while that in 
the Higgs sector is only 0.060. These specific values are, of course, a
result of the parameters chosen in our initialization procedure. They
would be different if other parameter values were selected.

The larger total gauge energy per gauge degree of freedom 
compared to the total Higgs energy per Higgs degree of freedom
results in the higher temperature of the gauge sector. Equilibration 
between the two subsectors apparently takes much longer. Since both
subsectors are apparently equally ``chaotic'', the relative stability 
of the two temperatures of the subsystems indicates that energy
exchange {\it between} the gauge and Higgs sectors is suppressed. This 
will be discussed quantitatively in the following subsection.

%%%%%%%%%%%%%%%%%%%%%%%%%%%%%%%%%%%%%%%%%%%%%%%%%%%%%%%%%%%%%%%%%%%%%%%%%%%
\subsection{Long-time equilibration and relaxation time}
\label{subsec:long}
%%%%%%%%%%%%%%%%%%%%%%%%%%%%%%%%%%%%%%%%%%%%%%%%%%%%%%%%%%%%%%%%%%%%%%%%%%%

As seen in Fig.~\ref{F1}, the coupled SU(2) Yang-Mills-Higgs system shows 
a very rapid thermalization within each subsector, with the relaxation 
times being of the order of $1$ lattice unit. However, it turns out that 
the mutual equilibration between the two sectors is rather slow and 
requires much longer time. To study the long time behavior of the coupled 
system, we follow the evolution of the energies.

In Fig.~\ref{F2}, we plot as a function of time (on a logarithmic scale) 
the following energies: $E_{\rm el}$ (solid), $E_{\rm mag}$ 
(dotted), $E_{\rm G}$ (short dash), $E_{\rm H}$ (long dash), 
and $E_{\rm t}$ (dot-dash). Here $E_{\rm el}$ is the electric energy 
per gauge degree of freedom:
 \begin{equation}
   E_{\rm el} = \frac{1}{6N^3}H_{\rm el} \,,
 \end{equation}
$E_{\rm mag}$ the magnetic energy per gauge degree of freedom:
 \begin{equation}
   E_{\rm mag} = \frac{1}{6N^3}H_{\rm mag} \,,
 \end{equation}
$E_{\rm H}$ the total Higgs energy per Higgs degree of freedom:
 \begin{equation}
   E_{\rm H} = \frac{1}{4N^3}(H_{\rm kin} + H_{\rm pot}) \,,
 \end{equation}
$E_{\rm G}$ the total gauge energy per gauge degree of freedom:
 \begin{equation}
   E_{\rm G}=E_{\rm el}+E_{\rm mag} \,,
 \end{equation} 
and $E_{\rm t}$ the total energy per degree of freedom in the entire system:
 \begin{equation}
 \label{Et}
   E_{\rm t}=(6 E_{\rm G}+4 E_{\rm H})/10 \,. 
 \end{equation} 
$N^3$ is the number of lattice sites. Note that each lattice site is 
associated with $10$ degrees of freedom: $6$ gauge degrees of freedom 
and $4$ Higgs degrees of freedom. The parameters used in the calculation 
are the same as in Fig.~\ref{F1}. In the weak coupling regime, $E_{\rm H}$ 
is the temperature of the Higgs sector as $E_{\rm G}$ is that of
the gauge sector, while $E_{\rm t}$ will be the final temperature of 
the entire system after the coupled gauge and Higgs sectors reach complete 
equilibrium.

Again, we notice that the energy transfer from the magnetic sector to 
the initially unpopulated electric sector occurs on a time scale of the 
order of unity, which agrees with Fig.~\ref{F1}. The same has been seen 
for the Higgs kinetic energy and the Higgs potential energy (not shown here). 
However, it takes a very long time for the initially hotter gauge sector 
to transfer some of its energy to the cooler Higgs sector. As shown in 
Fig.~\ref{F2}, $E_{\rm G}$ (short dash) and $E_{\rm H}$ (long dash) 
approach the flat line $E_{\rm t}\approx 0.31$ (dot-dash) only after 
times of many thousand lattice units. This means that the mutual 
equilibration process is associated with a very long relaxation time. 

Assuming an exponential relaxation law for the gauge field energy 
$E_{\rm G}(t)$,
 \begin{equation}
   E_{\rm G}(t)=E_{\rm G}(\infty)+E_0 \exp(-t/\tau_{\rm GH}) \,,
 \end{equation}
with an adjustable constant $E_0$, we can determine the relaxation 
time $\tau_{\rm GH}$ for the mutual equilibration process from a linear
fit to $\ln[(E_{\rm G}(t)-E_{\rm G}(\infty))]$. In Fig.~\ref{F3} we 
plot $\ln[(E_{\rm G}(t)-E_{\rm G}(\infty))/E_{\rm G}(\infty)]$ as a 
function of time; $E_{\rm G}(\infty)$ is taken as $E_{\rm t}$ from 
Eq.~(\ref{Et}). The solid line corresponds to the actual data while 
the dotted line represents the linear fit. From its slope we extract
$\tau_{\rm GH}=6002\pm72$ lattice units.

As shown in (\ref{rate}), the rate of energy transfer from the gauge sector 
to the Higgs sector is given by the sum $\sum E_{x,i}^a \, j_{x,i}^a$. 
In a large system with no correlations between the signs of $E_{x,i}^a$ 
and $j_{x,i}^a$, this sum would be a fluctuating quantity with a zero 
mean value. In our calculations, the sum turns out to be fluctuating 
around a small negative value, which means on average there is energy 
flowing from the gauge sector to the Higgs sector although the flow is 
very slow. Furthermore, from (\ref{EUdot}) and (\ref{hj}), we notice 
that $\sum E_{x,i}^a \, j_{x,i}^a$ is initially proportional to $\delta$ 
and to $R^2$. Both are restricted to small values if we want to work 
both in the weak coupling regime and close to the continuum limit. For 
larger couplings (i.e. for increasing $R$), keeping $\delta$ small to 
preserve the proximity to the continuum limit, one expects faster 
mutual equilibration. This is borne out numerically: increasing $R$ by a
factor 10 at fixed $\delta$, we found a much shorter relaxation time 
$\tau_{\rm GH}\approx 100$. This is, however, still much longer than the 
time scales for the thermalization of the gauge and the Higgs fields 
individually, which did not change appreciably compared to the case 
with the smaller value for $R$.

%%%%%%%%%%%%%%%%%%%%%%%%%%%%%%%%%%%%%%%%%%%%%%%%%%%%%%%%%%%%%%%%%%%%%%%%%%%
\section{Determination of the Lyapunov exponents}  
\label{sec:lyap}
%%%%%%%%%%%%%%%%%%%%%%%%%%%%%%%%%%%%%%%%%%%%%%%%%%%%%%%%%%%%%%%%%%%%%%%%%%%

Our main goal in this Section is to determine from the numerical integration 
of the equations of motion the Lyapunov exponents associated with the 
gauge and the Higgs fields, and to relate them to the damping rates 
obtained in perturbative calculations at finite temperature \cite{BT96}.

Let us begin with the discussion of two technical issues. First, when
determining the Lyapunov exponent from the exponential growth of the
distance between two initially nearby configurations, one can in 
principle use different distance measures. However, for all possible 
choices the exponential growth in the distance will be dominated by 
the eigenmode associated with the largest Lyapunov exponent, which is 
unique to a dynamical system. Unless the chosen distance measure accidentally
projects out this most unstable mode, different distance measures should
thus yield the same Lyapunov exponent. (Depending on the power with 
which the classical fields appear in the distance measure the actually 
measured growth rates can be different, but they are always related to 
the Lyapunov exponent by a simple multiplicative constant.) Based on these
remarks one expects the two distance measures given in (\ref{Gdist}) 
and (\ref{Hdist}) to yield the same Lyapunov exponent.

Second, in order to relate results from numerical integration to those 
from perturbative calculations, we need to study the scaling behavior 
of the Lyapunov exponents with the energy (temperature) of the system. 
For the pure SU(2) gauge system the relation
 \begin{equation}
   \lambda_{\rm G} a \approx {1\over 6} g^2 E_{\rm p} a 
 \label{oneover6}
 \end{equation}
was established \cite{Bi94}, with $E_{\rm p} = 6 E_{\rm G} /3$ denoting 
the gauge energy per plaquette. In \cite{BT96} it was shown by a perturbative
calculation with hard thermal loop resummation that, at leading order in 
the coupling constant, in the coupled SU(2) Yang-Mills-Higgs system 
the damping rate of a static gauge boson is the same as that in pure 
SU(2) gauge theory:
 \begin{equation}
   \gamma_{\rm G}(0)=0.176 \,g^2 \,T \,.  
 \label{scaling_1}
 \end{equation}
For the Higgs damping rate the relation 
 \begin{equation}
   \gamma_{\rm H}(0) = 0.018 \, g^2 \, T 
 \label{higgsdamp}
 \end{equation}
was found \cite{BT96} to hold for small values of the Higgs self 
coupling $\bar \lambda$. 

Using the scaling law (\ref{oneover6}) together with the fact that
in weak coupling the energy per plaquette $E_{\rm p}$ can be linearly
related to the temperature of the system, the authors of \cite{Bi94,Bi95}
found that in the pure SU(2) and SU(3) gauge systems the scaled largest 
Lyapunov exponent coincides with twice the gauge boson damping rate. 
What should one expect in our case? In Ref.~\cite{BT96} 
it was speculated that here the gauge boson damping rate should be related 
with the Lyapunov exponent extracted from the gauge distance measure 
(\ref{Gdist}) while the Higgs boson damping rate could be connected with 
the Lyapunov exponent extracted from the Higgs distance measure 
(\ref{Hdist}). Since the numerical results did not support this 
expectation, the authors questioned the existence of a simple relationship 
between damping rates and Lyapunov exponents in general, raising the
possibility that the coincidence found in pure SU(2) and SU(3) gauge 
theory was a numerical accident.

Since, however, the largest Lyapunov exponent is related to the most 
unstable mode in the entire coupled system and thus, as discussed above, is 
picked up by essentially every distance measure one can choose, we argue 
that one should rather expect both measures to yield the same Lyapunov 
exponent. We will show promptly that the numerical computations confirm 
this expectation. If this is the case the Lyapunov exponent can be related 
with at most one of the two damping rates. Since the gluon damping rate 
(\ref{scaling_1}) is much larger than the Higgs damping rate (\ref{higgsdamp})
it is clear that if there is such a relation it should be between the 
{\em gluon} damping rate and the maximal Lyapunov exponent. To which 
classically computable observable the smaller Higgs damping rate can be
related remains at this moment unclear. Of course, it could also be that the 
Lyapunov exponent is related to neither of the damping rates. To clarify 
this issue a careful study of the energy scaling of the Lyapunov exponent
is indispensible.

In the upper part of Fig.~\ref{F4} we show that, within the numerical 
accuracy, the two distance measures (\ref{Gdist}) and (\ref{Hdist}) 
indeed yield the same Lyapunov exponent. (Please note the suppressed 
zero on the vertical axis.) Thus they both measure the largest Lyapunov 
exponent of the entire coupled system. We also see that the Lyapunov
exponent shows a slight, but characteristic long-time variation which 
is the same for both distance measures. This behaviour will be further 
discussed below.

To understand the magnitude and physical meaning of the extracted
Lyapunov exponent, we studied its scaling behaviour with the
energy contained in the system. To this end we varied the relative 
energy contents in the gauge and Higgs subsectors as well as the total 
energy of the system, by changing $\delta$ and $R$ within the permitted 
ranges. We found that the Lyapunov exponent (which reflects the most 
unstable mode in the {\em entire} coupled system) does not scale with 
the total energy of the entire system, but rather with the energy stored 
in the gauge subsector. These scaling tests were performed on the basis
of numerical runs over a few hundred to a few thousand lattice time units. 
On these time scales the scaling of the largest Lyapunov with the gauge 
field energy per plaquette was found to be accurate on the level of 5 -- 10\%.
(Note that in Fig.~\ref{F4} the Lyapunov exponent shows long-time variations
over a similar range of about 10\%.) These findings indicate that the largest 
Lyapunov exponent is associated which the gauge field dynamics, and that 
the chaoticity in the gauge sector dominates the dynamics of the entire 
system, including the equipartition between potential and kinetic 
energy contributions in the Higgs subsystem.

Returning to Fig.~\ref{F4} we can now discuss the magnitude of the 
Lyapunov exponent scaled by the energy per plaquette $E_{\rm p}$ in the
gauge field sector. The lower part in Fig.~\ref{F4} shows that for very 
large times of order ten thousand lattice units the Lyapunov
exponent seems to converge to a constant value slightly above 0.16.
This is close to the value of about 1/6 which, according to the findings
of Ref.~\cite{Bi94}, is expected (see Eq.~(\ref{oneover6})) if the 
Lyapunov exponent is related to the gluon damping rate. The 
observed long-time variation of the Lyapunov exponent with final 
convergence only after many thousand time steps is apparently connected
with the already observed long-time variation of the relative energy 
content of the Higgs and gauge field subsectors seen in Fig.~\ref{F2}.
As the gauge and Higgs field sectors finally reach mutual equilibrium after
several ten thousand lattice time units, the scaled Lyapunov exponent 
$\lambda_{\rm G}/E_{\rm p}$ settles down to a constant asymptotic value 
of approximately 1/6. Noting that $E_{\rm p}=2 T$ after full equilibration
and inserting $g=1$, $a=1$, we obtain
 \begin{equation}
   \frac{\lambda_{\rm G} a}{g^2 T a} \approx 0.32  \,,  
 \label{scaling_2}
 \end{equation}
which together with (\ref{scaling_1}) verifies (\ref{scaling}) at the level
of 10\%. A more accurate verification is prohibited by the extremely
slow equilibration between the gauge and Higgs subsectors in our coupled 
system. The numerically determined Lyapunov exponent is somewhat on the 
low side; according to Figs.~\ref{F4} and \ref{F2}, the magnitude of
this discrepancy is directly related to the deviation of the gauge and 
Higgs field energies from equipartition.

Let us summarize the findings from this Section: We determined the gauge 
field Lyapunov exponent and verified that for the Yang-Mills-Higgs system
it is related to the static gauge boson damping rate in the same way as for
the pure gauge theory. The quantity $\lambda_{\rm H}$, as defined in 
(\ref{Hdist}) and (\ref{lyap_def}), was shown to give the same Lyapunov 
exponent and therefore to be governed by the gauge degrees of 
freedom, too. Therefore $\lambda_{\rm H}$ cannot be related to any
specific dynamics of the Higgs field, and it should not be called
``Higgs Lyapunov exponent'' as suggested in \cite{BT96}. In particular,
it is not possible to establish a relation between $\lambda_{\rm H}$ and
the Higgs damping rate (\ref{higgsdamp}). Unfortunately, we have not been 
able to come up with an alternative observable in the dynamics of
classical fields which could be related to $\gamma_{\rm H}$. 

%%%%%%%%%%%%%%%%%%%%%%%%%%%%%%%%%%%%%%%%%%%%%%%%%%%%%%%%%%%%%%%%%%%%%%%%%%%
\section{Further discussion}  
\label{sec:sum}
%%%%%%%%%%%%%%%%%%%%%%%%%%%%%%%%%%%%%%%%%%%%%%%%%%%%%%%%%%%%%%%%%%%%%%%%%%%

We close this paper with a discussion of our results in comparison with
the recent work presented in \cite{BT96}. For small values of the Higgs 
self coupling $\bar \lambda$, Bir\'o and Thoma \cite{BT96} found that 
the Lyapunov exponents $\lambda_{\rm H}$ and $\lambda_{\rm G}$ deduced 
from the Higgs and gauge fields, respectively, agree within the 
numerical accuracy. For very large values of the Higgs self coupling, 
on the other hand, $\lambda_{\rm H}$ was found to decrease, 
thereby breaking the coincidence between $\lambda_{\rm H}$ and 
$\lambda_{\rm G}$. For the case of small $\bar\lambda$ we have verified 
the coincidence of $\lambda_{\rm H}$ and $\lambda_{\rm G}$. It has
a natural explanation since, as we already discussed, in principle
any distance measure between two neighboring trajectories in phase 
space simply yields the maximal Lyapunov exponent of the entire system.

In view of this argument, the disagreement between $\lambda_{\rm H}$ and 
$\lambda_{\rm G}$ at large $\bar\lambda$ at first seems surprising.
If, however, the distance measure projects out the eigenmode with the
largest Lyapunov exponent, it can yield a different value for the 
growth rate. This is apparently what happens at very large values of 
$\bar\lambda$ for the Higgs distance (\ref{gidist}) used in \cite{BT96}: 
As already pointed out in \cite{BT96}, for large $\bar\lambda$ the dynamics 
of the Higgs amplitude (\ref{defHiggsamp}) becomes completely determined 
by the quartic self-interaction term (\ref{Elambda}), thereby 
effectively freezing the Higgs field amplitude $R$. Thus this 
degree of freedom starts to decouple from the rest of the system, and it may 
happen that the growth of perturbations in the Higgs amplitude is no longer 
influenced by the maximal Lyapunov exponent associated with the gauge field,
at least, over the time scale used for the determination of the Lyapunov
exponent.

We have not studied here the case of large $\bar\lambda$, for the following 
reasons: First, if one increases $\bar\lambda$ without decreasing the Higgs
amplitude one leaves the perturbative regime (\ref{weakcoupH}). Then 
the computed Lyapunov exponents can no longer be compared with the 
damping rates from hard thermal loop calculations. Second, for very large 
$\bar\lambda$ the energy density of the quartic interaction term 
(\ref{Elambda}) and thus also the total energy density become very large. 
As discussed in Sec.~\ref{subsec:iwc}, in this case one runs into lattice
artifacts due to the compactness of the magnetic energy on the lattice. 

While we agree with Bir\'o and Thoma \cite{BT96} about the coincidence 
of $\lambda_{\rm G}$ and $\lambda_{\rm H}$, we disagree with their 
determination of the temperature. Consequently we also differ in the 
interpretation of their results with respect to their connection with 
perturbative damping rates, believing that their rather negative conclusions
are not justified. Bir\'o and Thoma \cite{BT96} use the Stefan-Boltzmann 
law for 10 degrees of freedom
 \begin{equation}
   {E_{\rm tot} \over V}=
   {\pi^2 \over 3} k_{\rm B}T \left(\frac{k_{\rm B}T}{\hbar c}\right)^3
 \label{stefbol}
 \end{equation}
to extract the temperature $T$ from the total energy density 
$E_{{\rm tot}} / V$ of the system. Here we put in $k_{\rm B}$, $\hbar$,
and $c$ explicitly (elsewhere we take $k_{\rm B}=\hbar=c=1$).
As can be seen from the occurrence of
Planck's constant in (\ref{stefbol}), this law makes explicit use of the
quantized nature of excitation energies in a gas of (massless) particles. 
As such it can not be applied to a classical field theory. The classical 
Yang-Mills-Higgs fields on the lattice correspond to a dynamical system of 
coupled anharmonic oscillators which in thermal equilibrium and for weak 
coupling obeys the relation
 \begin{equation}
   {E_{\rm tot} \over N^3 a^3} = {10 T \over a^3} + o(g,\lambda)  \,. 
 \label{harmos}
 \end{equation} 
Thus the energy per degree of freedom is directly proportional to the 
temperature, and not to its fourth power. 

Even more crucial is our finding that the whole system equilibrates only 
after a very long time. Since the authors of \cite{BT96} did not use 
the rescaling method for extracting the Lyapunov exponents, they ran 
very soon into the problem of distance saturation due to the compactness
of the gauge group \cite{Bi94}, and thus their analysis was necessarily 
restricted to times which were short compared to the mutual equilibration 
time between the gauge and Higgs sectors. If the Lyapunov exponent is 
determined on such short time scales, the total energy of the system 
does not tell anything about the temperature of the relevant gauge 
subsector no matter whether it is determined from (\ref{stefbol}) or 
(\ref{harmos}). For this reason Bir\'o and Thoma missed the specific 
connection of the Lyapunov exponent with the gauge subsystem rather than
with the entire coupled Yang-Mills-Higgs system. Only an analysis 
like the one presented here can uncover the fact that after very
short thermalization times one can cleanly define separate temperatures 
for the gauge and the Higgs subsystems, and that the maximal Lyapunov 
exponent actually scales with the temperature of the gauge sector already
long before global equilibration between the gauge and Higgs sectors
sets in. 

In conclusion, we have studied the high temperature behavior of the coupled 
Yang-Mills-Higgs system in the weak coupling regime close to the continuum 
limit. We found that the coupled system undergoes thermalization in stages:
while the gauge and Higgs subsectors self-thermalize very rapidly, 
equipartition of the energy between these two sectors takes several 
orders of magnitude longer. We expect and have qualitative evidence for 
a connection between the rapid thermalization time scale within each 
of the two subsectors and the magnitude of the largest Lyapunov exponent;
to establish this relation quantitatively requires, however, a much more 
detailed and computer-intensive investigation. Since we found that the 
largest Lyapunov exponent is associated with the gauge field and scales with 
the gauge field energy, we conclude that thermalization of both the gauge 
and Higgs subsectors is driven by the chaotic dynamics of classical 
non-Abelian gauge fields. The Higgs subsector thus thermalizes via its
coupling to the chaotic gauge fields (gluon heat bath) and not due
its own intrinsic non-linearity via the selfcoupling $\bar\lambda$.
The scaling of the Lyapunov exponent with the gauge field energy
was used to verify that, even in the presence of a coupling to matter 
fields, the largest Lyapunov exponent still agrees with twice the damping 
rate of a static gauge boson. 

The analytical results of Thoma and Bir\'o \cite{BT96} from thermal
perturbation theory show that in the coupled Yang-Mills-Higgs system 
also the Higgs damping rate is proportional to $g^2 T$ and thus, from 
the general arguments given e.g. in \cite{Bi95}, is expected to survive
in the classical limit. At the moment it remains an open question whether
it is possible to identify an observable in the dynamical evolution
of classical Yang-Mills-Higgs fields which can be associated with the 
Higgs boson damping rate in a similar way as the largest Lyapunov 
exponent is related to the gluon damping rate.

%%%%%%%%%%%%%%%%%%%%%%%%%%%%%%%%%%%%%%%%%%%%%%%%%%%%%%%%%%%%%%%%%%%%%%%%%%%%%
\acknowledgements
%%%%%%%%%%%%%%%%%%%%%%%%%%%%%%%%%%%%%%%%%%%%%%%%%%%%%%%%%%%%%%%%%%%%%%%%%%%%%

The authors thank M.~Thoma and T.~Bir\'o for fruitful discussions.
This work was supported in part by the U.S. Department of Energy 
(Grant No. DE-FG02-96ER40945) and by the North Carolina Supercomputing Center.
U.H. and S.L. acknowledge support by the Deutsche Forschungsgemeinschaft 
(DFG) and the Bundesministerium f\"ur Bildung, Wissenschaft, Forschung und
Technologie (BMBF). 

%%%%%%%%%%%%%%%%%%%%%%%%%%%%%%%%%%%%%%%%%%%%%%%%%%%%%%%%%%%%%%%%%%%%%%%%%%%%%
\appendix
\section{Calculation of phase space factors}
\label{sec:appendix} 
%%%%%%%%%%%%%%%%%%%%%%%%%%%%%%%%%%%%%%%%%%%%%%%%%%%%%%%%%%%%%%%%%%%%%%%%%%%%%

In this Appendix we derive the phase space factors presented in 
Sec.~\ref{subsec:short}. In general it is difficult to calculate the energy 
distribution over the microscopic degrees of freedom of an arbitrary system 
evolving in time. It is equivalent to solving the equations of motion. 
However, as shown in the main part of the paper, the gauge and Higgs 
sectors of the system equilibrate separately very rapidly (albeit 
reaching different temperatures). In addition we are working in a 
regime where weak coupling is assumed between the gauge and Higgs 
sectors as well as within each sector. This significantly simplifies the 
task of calculating the energy distributions of different degrees of freedom. 

To determine how the electric energy is distributed over the lattice sites 
for the thermalized gauge sector at temperature $T_{\rm G}$, we neglect 
its coupling to the Higgs fields and write the partition function for
the gauge sector as
 \begin{equation}
   Z = \int \prod\limits_{x,i,a} d\bar E_{x,i}^a \, d\mu(U_{x,i}) \, 
   \exp\left[- {\bar H_{\rm el}+ \bar H_{\rm mag} \over T_{\rm G}} \right] \, 
   \delta(D_{x,i}^{ab} \bar E_{x,i}^b) \,  \delta(F(U_{x,i})) \,,   
 \label{partition}
 \end{equation}
where the first $\delta$-function takes into account the constraints imposed
by Gauss' law,  and the second one represents gauge fixing of the
form $F(U_{x,i})=0$, which involves link variables only. $\mu(U)$ is the Haar 
measure defined on the SU($2$) group mainfold. 

For the electric energy on a site, we define its distribution function 
$P_{\rm el}(\epsilon)$ as
 \begin{eqnarray}
   P_{\rm el}(\epsilon) & = &  Z^{-1}\, 
   \int \prod\limits_{x,i,a} d\bar E_{x,i}^a \, d\mu(U_{x,i}) \, 
   \exp\left[- {\bar H_{\rm el}+ \bar H_{\rm mag} \over T_{\rm G}} \right] \,
 \nonumber \\
   && 
   \times\,\delta(D_{x,i}^{ab} \bar E_{x,i}^b) \, \delta(F(U_{x,i})) \, 
   \delta(\epsilon - {\textstyle {1\over 2}} \sum\limits_{i,a}
   \bar E_{x_0,i}^a \bar E_{x_0,i}^a)
 \nonumber \\
   &=& Z^{-1}\,
   \int \prod\limits_{x,i,a} d\bar E_{x,i}^a \, d\mu(U_{x,i}) \, 
   d\lambda_x^a \, d\kappa \, \delta(F(U_{x,i}))
 \nonumber \\ 
   && \times 
   \exp\left[- {\bar H_{\rm el}+\bar H_{\rm mag} \over T_{\rm G}} 
             + i \sum\limits_{x,i,a,b} \lambda_x^a D_{x,i}^{ab} \bar E_{x,i}^b 
             + i \kappa \, (\epsilon - {\textstyle {1\over 2}}
               \sum\limits_{i,a} \bar E_{x_0,i}^a \bar E_{x_0,i}^a) 
       \right]  \,,
 \label{Pel}
 \end{eqnarray}
where we have introduced Lagrange multipliers $\lambda_x^a$ and $\kappa$ to
account for the $\delta$-functions. The electric and magnetic energies 
are defined in (\ref{Hel}) and (\ref{Hmag}), respectively. Since the 
electric fields appear only up to quadratic order in (\ref{Pel}), the 
integrations over $\bar E_{x,i}^a$ can be performed with the result
 \begin{eqnarray}
   P_{\rm el}(\epsilon) & \sim & \int \prod\limits_{x,i,a} d\mu(U_{x,i}) \, 
   d\lambda_x^a \, d\kappa \, \delta(F(U_{x,i})) \, 
   \det{}^{-1/2} \left[ \left( {1\over T_{\rm G}} 
                               + i \kappa \delta_{x,x_0}\right) 
                        \delta_{x,x'}\,\delta_{i,i'}\,\delta_{a,a'} \right] 
 \nonumber \\
   && \times 
   \exp\left[- {\bar H_{\rm mag} \over T_{\rm G}} 
             - {T_{\rm G} \over 2} \sum\limits_{x,i,a,b}^{ x \neq x_0} 
               (D_{x,i}^{ab} \lambda_x^a)^2 
             - {T_{\rm G}\over {2(1+i\kappa T_{\rm G})}}
               \sum\limits_{i,a,b} (D_{x_0,i}^{ab} \lambda_{x_0}^a)^2
             + i \kappa \epsilon 
       \right]  
 \nonumber \\
   & \sim & \int \prod\limits_{x,i,a} d\mu(U_{x,i}) \, 
   d\lambda_x^a \, d\kappa \, \delta(F(U_{x,i})) \, 
   \left( 1 +i \kappa T_{\rm G}\right)^{-9/2} 
 \nonumber \\
    && \times 
    \exp\left[- {\bar H_{\rm mag} \over T_{\rm G}} 
              - {T_{\rm G} \over 2} \sum\limits_{x,i,a,b} 
                {{(D_{x,i}^{ab} \lambda_x^a)^2}
                 \over{1+i\kappa T_{\rm G}\delta_{x,x_0}}} 
              + i \kappa \epsilon 
        \right]  \,.
 \end{eqnarray}
The integration over the Lagrange multipliers $\lambda_x^a$ can be evaluated
similarly. After that, the whole integral factorizes into integrations
over the link variables and the Lagrange multiplier $\kappa$.
The integration over the link variables is $\epsilon$-independent and 
can be factored out as a constant. We thus obtain 
 \begin{eqnarray}
   P_{\rm el}(\epsilon) 
   & \sim & \int\limits^{+\infty}_{-\infty} d\kappa \, 
     \left( 1 +i \kappa T_{\rm G} \right)^{-3} 
     \exp\left( i \kappa \epsilon \right)  
 \nonumber \\ 
   & \sim & \frac{\epsilon^2}{T^3_{\rm G}} \exp(-\epsilon/T_{\rm G}) \,,
 \label{Pelres}
 \end{eqnarray}
which yields the electric phase space factor $f_{\rm el}(\epsilon) = 
\epsilon^2$. Three points should be stressed here: First, the above 
result is exact for a pure SU(2) gauge system and approximately valid 
for the weakly coupled SU(2) Higgs system. Second, the apparent 
temperature agrees with the true temperature, i.e.~$C_{\rm el} =1$. 
Finally we mention that $\epsilon^2$ is replaced by $\epsilon^{N^2-2}$ 
for the general case of SU($N$). 

In deriving the distribution for the magnetic energy on a plaquette, we again 
neglect the coupling to the Higgs fields and write
 \begin{eqnarray}
   P_{\rm mag}(\epsilon) &=& 
   Z^{-1}
   \int \prod\limits_{x,i,a} d\bar E_{x,i}^a \, d\mu(U_{x,i}) \, 
   \exp\left[- {\bar H_{\rm el}+\bar H_{\rm mag} \over T_{\rm G}} \right] \,
 \nonumber   \\
   &&
   \times\,\delta(D_{x,i}^{ab} \bar E_{x,i}^b) \,
   \delta(F(U_{x,i})) \,  
   \delta\left( \epsilon - 4 (1 - {\textstyle {1\over 2}}{\rm tr} \, 
                U_{x_0,i_0\,j_0})
         \right) \,.
 \end{eqnarray}

Assuming weak coupling within the gauge sector itself, we neglect the nonlinear
term in the Gauss' law, which reduces ${\bf D}\cdot{\bf E}=0$ to a
linear form: $\nabla\cdot{\bf E}=0$. Then the integration over the electric
fields decouples from that over the link variables and can be factored out as a
constant independent of $\epsilon$. Hence we find
 \begin{equation}
   P_{\rm mag}(\epsilon) \sim
   \int \prod\limits_{x,i} d\mu(U_{x,i}) \, 
   \exp\left(- {\bar H_{\rm mag}/T_{\rm G}} \right) \,
   \delta(F(U_{x,i})) \,  
   \delta\left(
   \epsilon - 4 (1 - {\textstyle {1\over 2}}{\rm tr} \, U_{x_0,i_0\,j_0})
   \right) \,.
 \end{equation}
Now the integration is over the link variables $U_{x,i}$, while the energy is 
expressed in terms of the plaquettes $U_{x,ij}$. In order to carry out 
the integration, we use a technique described in \cite{Batrouni82} to 
transform the integral over link variables into an integral over plaquettes 
\cite{Thesis94}: 
 \begin{equation}
   P_{\rm mag}(\epsilon) 
   \sim
   \int \prod\limits_{x,i,j} d\mu(U_{x,ij})  \, 
   \exp\left(- {\bar H_{\rm mag}/T_{\rm G}} \right) \, 
   \prod\limits_c\delta(U_c-1) \, 
   \delta\left(
   \epsilon - 4 (1 - {\textstyle {1\over 2}}{\rm tr} \, U_{x_0,i_0\,j_0})
   \right) \,,   
 \label{Pmag0}
 \end{equation}
where the $\delta$-function $\delta(U_c-1)$ imposes a constraint on each
elementary cube $c$. These constraints are the lattice version of the 
Bianchi identities, which correspond to ${\bf D}\cdot{\bf B}=0$ in the 
continuum limit. $U_c$ is an SU($2$) matrix defined in terms of the link 
variables associated with the cube $c$. The transformation from links to 
plaquettes corresponds to the transformation from vector potentials to 
magnetic field strengths in the continuum limit. For details on the 
transformation and the exact form of $U_c$, we refer the reader to 
\cite{Batrouni82} and \cite{Thesis94}. Now we consider (\ref{Pmag0}) 
in two limiting cases: $T_{\rm G}\rightarrow 0$ and $T_{\rm G}
\rightarrow\infty$. 

In the low temperature limit, i.e.~when the field amplitudes are small, 
the non-Abelian part of the constraints becomes negligible and hence 
the non-Abelian Bianchi identities ${\bf D}\cdot{\bf B}=0$ reduce to 
their Abelian equivalents $\nabla\cdot{\bf B}=0$. On the lattice, they 
take the form of $\sum_{ij\in c} B_{ij}=0$, where the sum runs through the
six plaquettes forming the surface of a cube. These are strong constraints 
which completely eliminate the longitudinal components of the $B$ fields 
and reduce the effective number of degrees of freedom by $1/3$.
So in this limit, the apparent temperature $T_{\rm a}=2 T_{\rm G}/ 3$.

In the high temperature limit, we expand the constraints in terms of 
the irreducible representations of the group SU($2$) \cite{Cr83}:
 \begin{equation}
   \delta(U_c-1) = \sum\limits_{J} (2J+1)\, \chi_{J}(U_c)\, ,
   \qquad J = 0, ~{\textstyle {1\over2}}, ~1, ~\dots
 \end{equation}
where $\chi_{J}(U_c)$ is the trace of $U_{c}$ in the $(2J+1)$-dimensional 
representation. For $T_{\rm G}\rightarrow\infty$, the dominant contribution 
to (\ref{Pmag0}) comes from $J=0$. Taking $J=0$ and neglecting those 
with $J\neq 0$ is equivalent to ignoring the constraints imposed by the 
Bianchi identities and treating the plaquettes as independent variables. 
Then it is not surprising to find that the apparent temperature $T_{\rm a}$ 
obtained in measuring $P_{\rm mag}(\epsilon)$ coincides with the actual 
temperature $T_{\rm G}$.

While it is easy to do the integrations in the limiting cases, it is
difficult to calculate (\ref{Pmag0}) for intermediate temperatures. 
Noting that the major effect of the constraints $\delta(U_c-1)$ on 
the distribution function is to change its logarithmic slope from
the actual temperature $T_{\rm G}$ to the apparent temperature 
$T_{\rm a}$, we assume that these constraints can be effectively 
taken into account by replacing $T_{\rm G}$ by $T_{\rm a}$ in (\ref{Pmag0})
and write
 \begin{equation}
  P_{\rm mag}(\epsilon) 
  \sim
  \int \prod\limits_{x,i,j} d\mu(U_{x,ij})  \, 
  \exp\left(- {\bar H_{\rm mag}/T_{\rm a}} \right) \,  
  \delta\left(
  \epsilon - 4 (1 - {\textstyle {1\over 2}}{\rm tr} \, U_{x_0,i_0\,j_0})
  \right) \,.   
 \label{Pmag1}
 \end{equation}
Of course, the validity of this assumption is to be justified and the 
relation between $T_{\rm a}$ and $T_{\rm G}$ to be established by 
numerical calculations.

The evaluation of (\ref{Pmag1}) is straightforward. Making use of the 
quaternion representation $U=u^0+i\tau^a u^a$ for a plaquette $U$ and 
the explicit form of the Haar measure on the SU($2$) group manifold
 \begin{equation}
   d\mu(U)= 
   {1\over{\pi^2}} 
   {\delta\left({\textstyle {1\over 2}} {\rm tr}(U^{\dag} U) -1 \right)}\,
   du^0\,du^1\,du^2\,du^3 \,,
 \end{equation}
one can easily carry out the integration in (\ref{Pmag1}) and obtain
 \begin{equation}
   P_{\rm mag}(\epsilon) \sim
   f_{\rm mag}(\epsilon) \exp\left(-\epsilon/T_{\rm a}\right) \,, 
 \label{Pmag2}
 \end{equation}
where $f_{\rm mag}(\epsilon)=\sqrt{\epsilon(8-\epsilon)}$. The relation 
$T_{\rm a} = C(T_{\rm G}) \, T_{\rm G}$ can be determined numerically and 
$C(T_{\rm G})$ varies from $2/3$ to $1$ as $T_{\rm G}$ increases from zero
to infinity \cite{Bi94}.

Note that, for the sake of convenience, we have calculated the distribution 
of the electric energy on a site and that of the magnetic energy on a 
plaquette. Hence the apparent electric and magnetic distributions do 
not coincide even at low temperature where, in principle, one can treat 
the electric and magnetic energy in exactly the same way and arrive at 
the same distributions. However, as seen in Fig.~\ref{F1}, the different
results for $T_{\rm a}$ relate to the same value of $T_{\rm G}$ in the 
electric and magnetic sector.

To derive how the Higgs kinetic and potential energies are distributed 
over the lattice sites for the thermalized Higgs sector at temperature 
$T_{\rm H}$, we neglect both the gauge-Higgs coupling and the Higgs 
self-coupling. Then it is straightforward to derive
 \begin{equation}
   P_{\rm kin,pot}(\epsilon) \sim  {\epsilon\over{T^2_{\rm H}}} 
   \exp(-\epsilon/T_{\rm H}) \,,  
 \label{PHiggs}
 \end{equation}
from which we obtain $f_{\rm kin}(\epsilon)=f_{\rm pot,2}(\epsilon)=\epsilon$.

In principle, perturbative corrections to the results obtained in
(\ref{Pelres}), (\ref{Pmag2}), and (\ref{PHiggs}) can be calculated. 
For our case at hand, however, it turns out that the formulae derived 
above are in good agreement with the numerical results. This is not 
surprising because our numerical calculations are performed for weak coupling.

%%%%%%%%%%%%%%%%%%%%%%%%%%%%%%%%%%%%%%%%%%%%%%%%%%%%%%%%%%%%%%%%%%%%%%%%%%%%%%

%\end{thebibliography}
%%%%%%%%%%%%%%%%%%%%%%%%%%%%%%%%%%%%%%%%%%%%%%%%%%%%%%%%%%%%%%%%%%%%%%%%%%%%

\begin{figure}
\caption{Short-time equilibration inside the gauge and the
Higgs subsectors. 
Four characteristic temperatures are plotted as functions of time:
the temperature of the electric gauge sector $T_{\rm el}$ (solid line), 
the temperature of the magnetic gauge sector $T_{\rm mag}$ (dashed line),
the temperature of the kinetic Higgs sector $T_{\rm kin}$ (dotted line),
and the temperature of the quadratic part of the potential Higgs 
sector $T_{\rm pot,2}$ (dot-dashed line).
Parameters: $\delta=0.2$, $R=0.2$, $g=1$, $\lambda=1$, and $a=1$. 
}
\label{F1}
\end{figure}

\begin{figure}
\caption{Long-time equilibration between the gauge and the Higgs
subsectors. 
Five energies are plotted as functions of time:
the electric energy per gauge field degree of freedom $E_{\rm el}$ (solid line),
the magnetic energy per gauge field degree of freedom $E_{\rm mag}$ (dotted line),
the total gauge field energy $E_{\rm G}=E_{\rm el}+E_{\rm mag}$ per gauge
field degree of freedom (short dash), 
the Higgs field energy per Higgs field degree of freedom $E_{\rm H}$ 
(long dash), and the total energy per degree of freedom $E_{\rm t}$
(dot-dashed line). Parameters as in Fig.~\protect\ref{F1}.
}
\label{F2}
\end{figure}

\begin{figure}
\caption{Determination of the relaxation time $\tau_{\rm GH}$ associated 
with the long-time equilibration between the gauge and the Higgs subsectors. 
The solid represents the numerical data while the dotted line is a 
linear fit to the data.}
\label{F3}
\end{figure}

\begin{figure}
\caption{
Upper graph: coincidence of $\lambda_{\rm G}$ and $\lambda_{\rm H}$.
Lower graph: convergence of the gauge field Lyapunov exponent.
$E_{\rm p}$ is the gauge field energy per plaquette.
The fields were rescaled after each period $\Delta t=30$. 
}
\label{F4}
\end{figure}

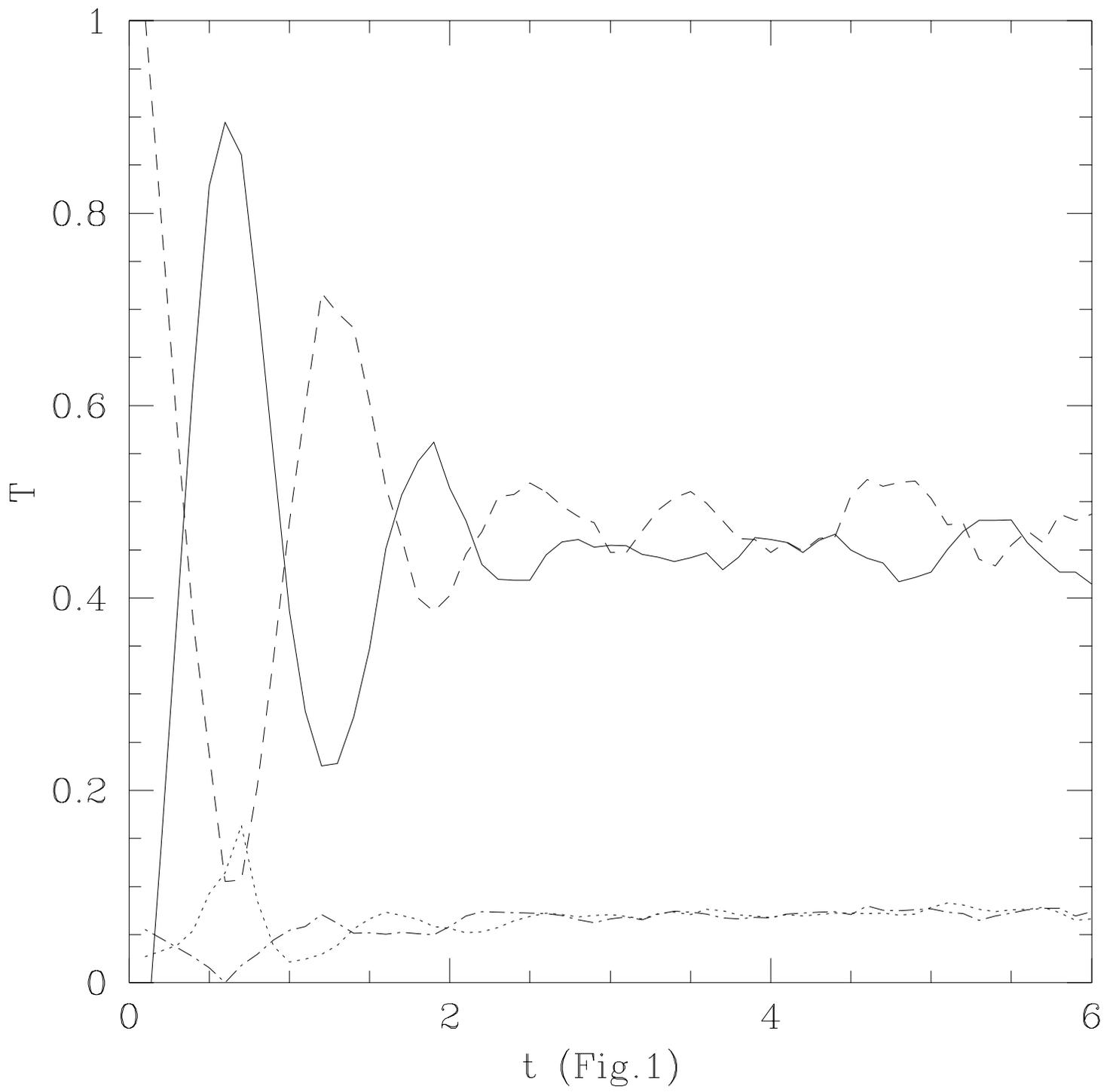
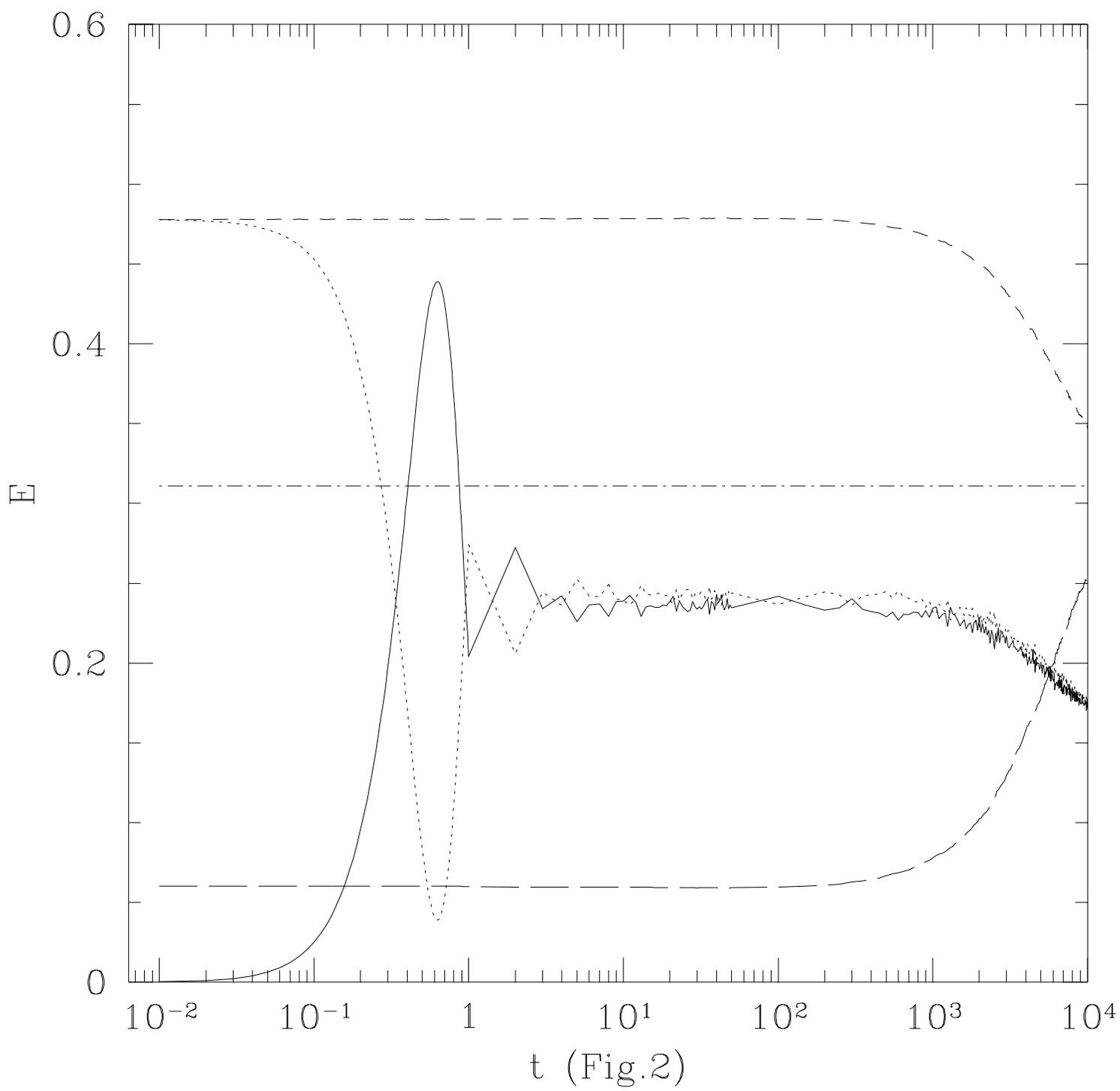
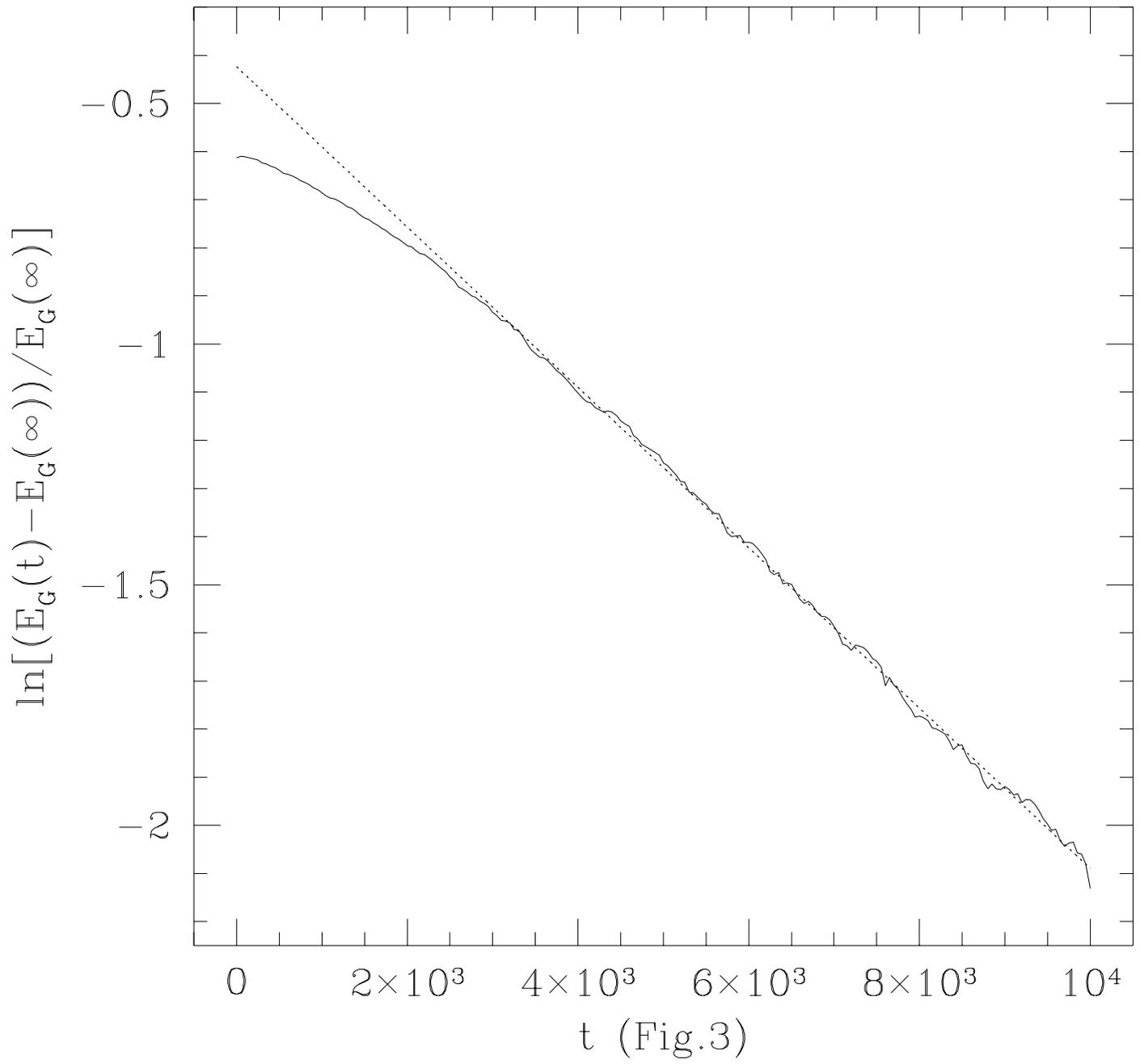
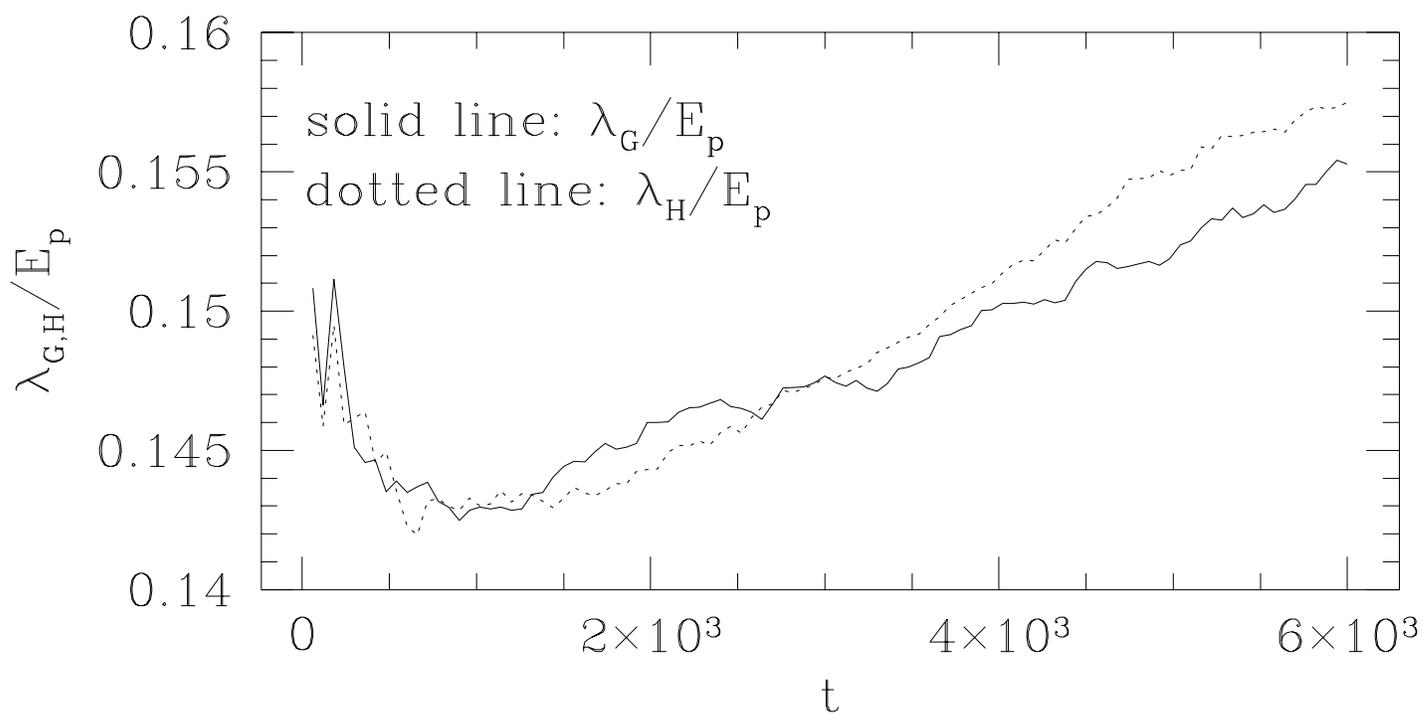
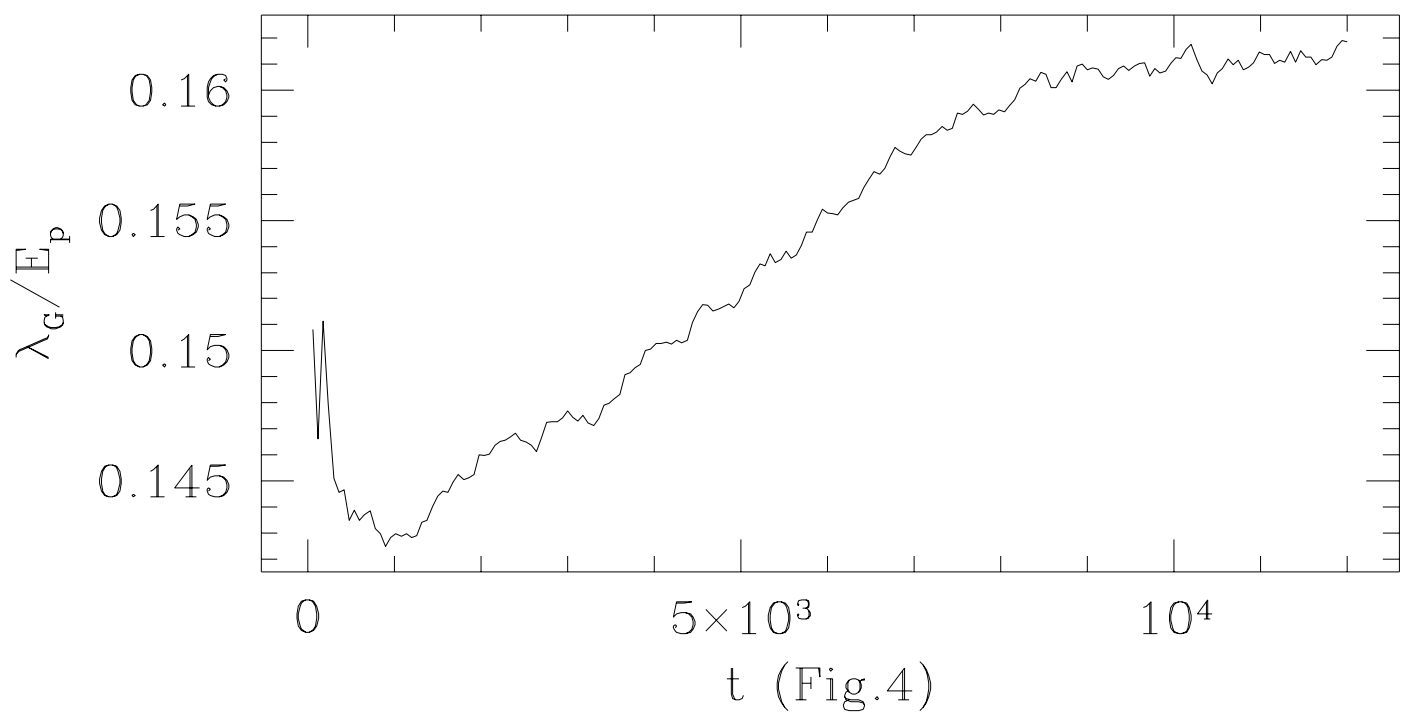
\end{document}